\documentclass[%
 reprint,
superscriptaddress,
 amsmath,amssymb,
 aps,
 prx,
]{revtex4-2}

\usepackage{graphicx}
\usepackage{dcolumn}
\usepackage{float}
\usepackage{braket}
\usepackage{SIunits}
\usepackage{comment}
\usepackage[usenames, dvipsnames]{color}
\usepackage[english]{babel}

\begin{document}


\title{Algorithmic Ground-state Cooling of Weakly-Coupled Oscillators using Quantum Logic} 



\author{Steven A. King}
\email[Contact: ]{steven.king@ptb.de}
\affiliation{Physikalisch-Technische Bundesanstalt, Bundesallee 100, 38116 Braunschweig, Germany}

\author{Lukas J. Spieß}
\affiliation{Physikalisch-Technische Bundesanstalt, Bundesallee 100, 38116 Braunschweig, Germany}

\author{Peter Micke}
\affiliation{Physikalisch-Technische Bundesanstalt, Bundesallee 100, 38116 Braunschweig, Germany}
\affiliation{Max-Planck-Institut f\"ur Kernphysik, Saupfercheckweg 1, 69117 Heidelberg, Germany}

\author{Alexander Wilzewski}
\affiliation{Physikalisch-Technische Bundesanstalt, Bundesallee 100, 38116 Braunschweig, Germany}

\author{Tobias Leopold}
\affiliation{Physikalisch-Technische Bundesanstalt, Bundesallee 100, 38116 Braunschweig, Germany}

\author{Jos\'e~R.~{Crespo~L\'opez-Urrutia}}
\affiliation{Max-Planck-Institut f\"ur Kernphysik, Saupfercheckweg 1, 69117 Heidelberg, Germany}

\author{Piet O. Schmidt}
\affiliation{Physikalisch-Technische Bundesanstalt, Bundesallee 100, 38116 Braunschweig, Germany}
\affiliation{Institut f\"ur Quantenoptik, Leibniz Universit\"at Hannover, Welfengarten 1, 30167 Hannover, Germany}

\date{\today}

\begin{abstract}
Most ions lack the fast, cycling transitions that are necessary for direct laser cooling. In most cases, they can still be cooled sympathetically through their Coulomb interaction with a second, coolable ion species confined in the same potential. If the charge-to-mass ratios of the two ion types are too mismatched, the cooling of certain motional degrees of freedom becomes difficult. This limits both the achievable fidelity of quantum gates and the spectroscopic accuracy. Here we introduce a novel algorithmic cooling protocol for transferring phonons from poorly- to efficiently-cooled modes. We demonstrate it experimentally by simultaneously bringing two motional modes of a Be$^{+}$-Ar$^{13+}$ mixed Coulomb crystal close to their zero-point energies, despite the weak coupling between the ions. We reach the lowest temperature reported for a highly charged ion, with a residual temperature of only $T\lesssim200$~\micro{K} in each of the two modes, corresponding to a residual mean motional phonon number of $\langle n \rangle \lesssim 0.4$. Combined with the lowest observed electric field noise in a radiofrequency ion trap, these values enable an optical clock based on a highly charged ion with fractional systematic uncertainty below the $10^{-18}$ level. Our scheme is also applicable to (anti-)protons, molecular ions, macroscopic charged particles, and other highly charged ion species, enabling reliable preparation of their motional quantum ground states in traps. 
\end{abstract}


\maketitle 


\section{Introduction}
Laser cooling has ushered in a new era of spectroscopic precision and accuracy. Atoms and ions can be brought practically to rest, greatly suppressing Doppler broadening and shifts. The required strong, cycling electronic transitions in the laser-accessible range only exist in a few species, however. This limitation can be overcome using so-called sympathetic cooling, whereby a second ion, referred to as the `cooling ion' (or `logic ion', depending on the application), is confined in the same trap together with the ion of interest, which we will refer to as the `spectroscopy ion'. Their mutual Coulomb interaction leads to coupled motion of the ions, which can be damped using the cooling ion \cite{larson_sympathetic_1986}. Applications to-date include quantum information processing \cite{kielpinski_sympathetic_2000,barrett_sympathetic_2003}, ultra-high-accuracy optical atomic clocks \cite{schmidt_spectroscopy_2005}, molecular ions \cite{blythe_production_2005, molhave_formation_2000, willitsch_chemical_2008, wan_efficient_2015, rugango_sympathetic_2015}, multiply charged ions \cite{Heugel_Yb2+}, highly charged ions (HCI) \cite{schmoger_coulomb_2015, micke_coherent_2020}, and helium ions \cite{roth_sympathetic_2005}.\par

The motional coupling between the ions depends on how well the charge-to-mass ratios of the two species match. At large mismatches, the ions can move almost independently in the trap. This reduces the efficiency of sympathetic cooling in modes of motion where the cooling ion is almost stationary \cite{wubbena_sympathetic_2012, home_chapter_2013, sosnova2020character}, thereby lengthening the cooling time, and raising the equilibrium temperature where heating mechanisms are present. Furthermore, the control and manipulation of these modes via the cooling ion (e.g., for sideband cooling or thermometry) is challenging. As such, these modes were expected to pose limitations to quantum protocols \cite{sosnova2020character} and achievable spectroscopic accuracy \cite{kozlov_hci}.\par

\begin{figure*}[ht!]
	\includegraphics[width=\textwidth]{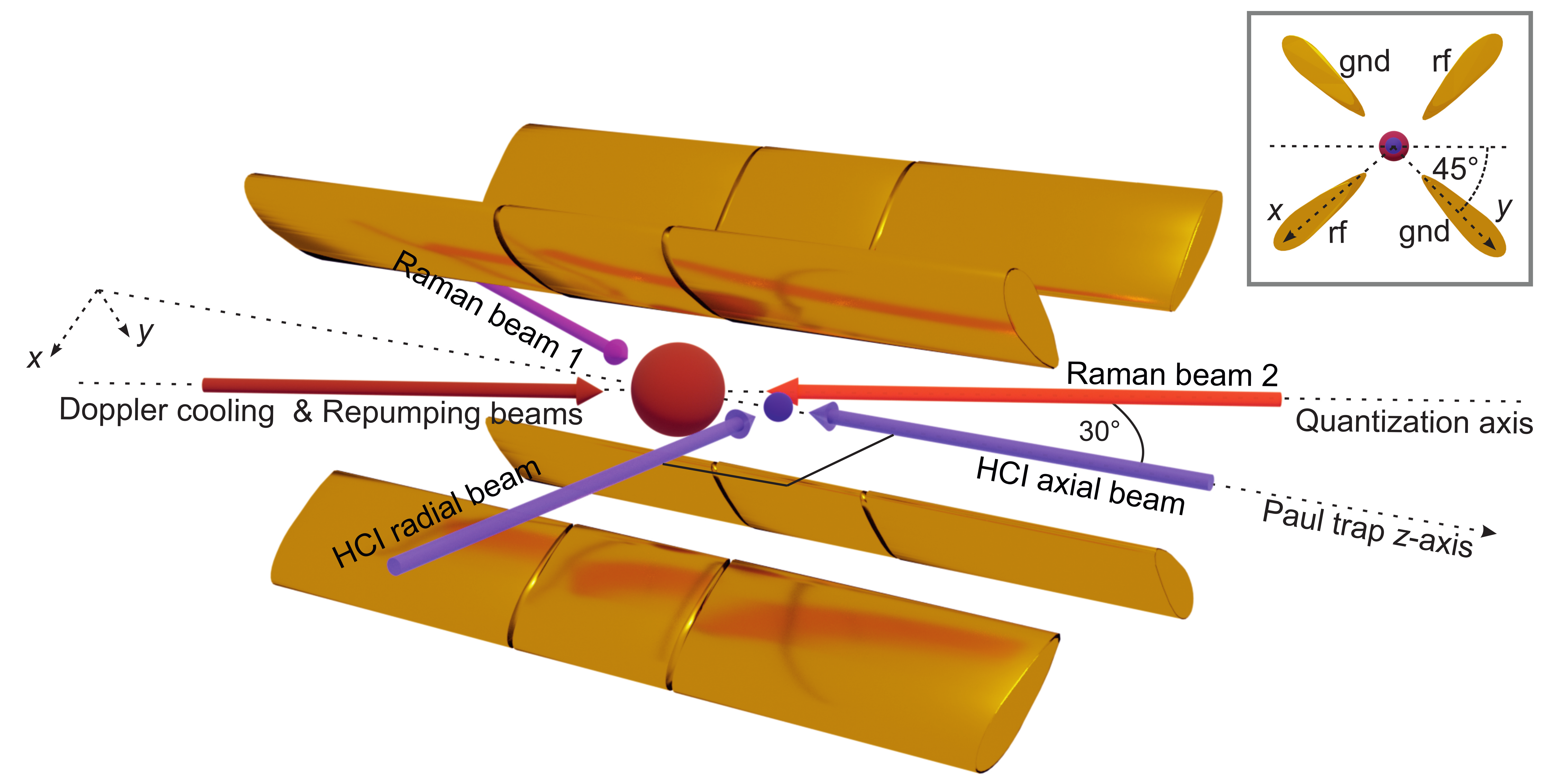}
	\caption{Simplified depiction (not to scale) of the two-ion Coulomb crystal of a single Be$^{+}$ ion (red, left) and a single HCI (purple, right) confined in a linear Paul trap. The laser beams needed for their manipulation are shown. The ions are located on the trap axis ($z$); radial motion evolves on the ($x$, $y$) coordinates defined by the trap electrodes which are at $45^{\circ}$ to the plane on which all the laser beams lie. The Doppler cooling, repumper, and Raman laser beams all intersect the $z$-axis at an angle of 30$^{\circ}$. The beam for addressing axial motion of the HCI is delivered along the $z$-axis, and the beam for addressing its radial motion is delivered perpendicular to it. Inset: radial cross section of the ion trap, showing the orientation of the radial trap axes as defined by the radiofrequency (rf) and ground (gnd) electrodes.}
	\label{fig:beams}
\end{figure*}

Several attempts were made to address this problem by coupling the weakly-cooled modes to others that are more effectively cooled. For example, in a linear radiofrequency (rf) Paul trap, a tilt was applied to a two-ion crystal previously aligned along the trap symmetry axis by means of a transverse electric field \cite{rosenband_frequency_2008}. An alternative approach was to mix the modes using additional radiofrequency fields \cite{gorman_two-mode_2014}. These techniques work well if the coupling is not too weak, but below a certain limit the induced mixing still cannot afford effective cooling.\par

In this work, we demonstrate how to remove phonons from weakly coupled modes using algorithmic cooling. It consists of a quantum protocol where coherent operations transfer entropy (or heat) from one part of a system to another, from which it can be removed by coupling, e.~g., to a bath \cite{boykin_algorithmic_2002, brassard_prospects_2014}. Originally, algorithmic cooling was developed to improve the polarization of a solid state spin sample without cooling the environment in nuclear magnetic resonance experiments \cite{baugh_experimental_2005}. Later, it was utilized to remove entropy from a quantum gas in an optical lattice \cite{popp_ground-state_2006, bakr_orbital_2011}; demon-like algorithmic quantum cooling was realized in a photonic quantum optical network \cite{xu_demon-like_2014}, and a partner-pairing algorithm was proposed to cool the motion of a single-species trapped ion quantum computer \cite{schulman_physical_2005}. Here, we demonstrate the technique by cooling weakly-coupled motional modes of a trapped, sympathetically-cooled highly charged ion. By using the internal spin of the HCI, excitation of the weakly-coupled mode is coherently mapped onto a different mode that can be cooled efficiently by the cooling ion. The technique we present is widely applicable and could be used to improve the performance of many sympathetically-cooled systems with mismatched charge-to-mass ratios between the particles.\par

\section{Experimental setup}
\label{sec:experiment}

Detailed descriptions of our setup can be found in references \cite{micke_heidelberg_2018, leopold_cryogenic_2019, micke_closed-cycle_2019, micke_coherent_2020, supplementary}. In short: a two-ion crystal composed of a single $^{9}$Be$^{+}$ ion and a single $^{40}$Ar$^{13+}$ ion is confined in a cryogenic linear rf Paul trap, which is driven at a frequency of $\Omega=2\pi\times24.0$~MHz. The blade electrodes that generate the rf trapping potential are inclined at $45^{\circ}$ to the horizontal plane in a four-fold symmetric pattern. We define the $z$ direction to be along the axial direction of the trap, with the two radial directions $x$ and $y$ lying perpendicular to it and along the axes of the blade electrodes, as illustrated in Fig.~\ref{fig:beams}.\par

Choosing the axial confinement to be weaker than the radial confinement and careful compensation of stray dc electric fields forces the ions to arrange themselves along the $z$ axis, where the rf field has a node. The ions can oscillate along the three axes either in-phase with one another as part of center-of-mass motion, or out-of-phase in so-called `stretching' or `rocking' modes, leading to six normal modes of motion in total \cite{james_quantum_1998, kielpinski_sympathetic_2000, wubbena_sympathetic_2012}. Under our typical trapping conditions, the eigenfrequencies $\omega / 2 \pi$ of these modes lie in the range of $1-5$~MHz.\par

\begin{figure*}[ht!]
    \centering
	\includegraphics[width=\textwidth]{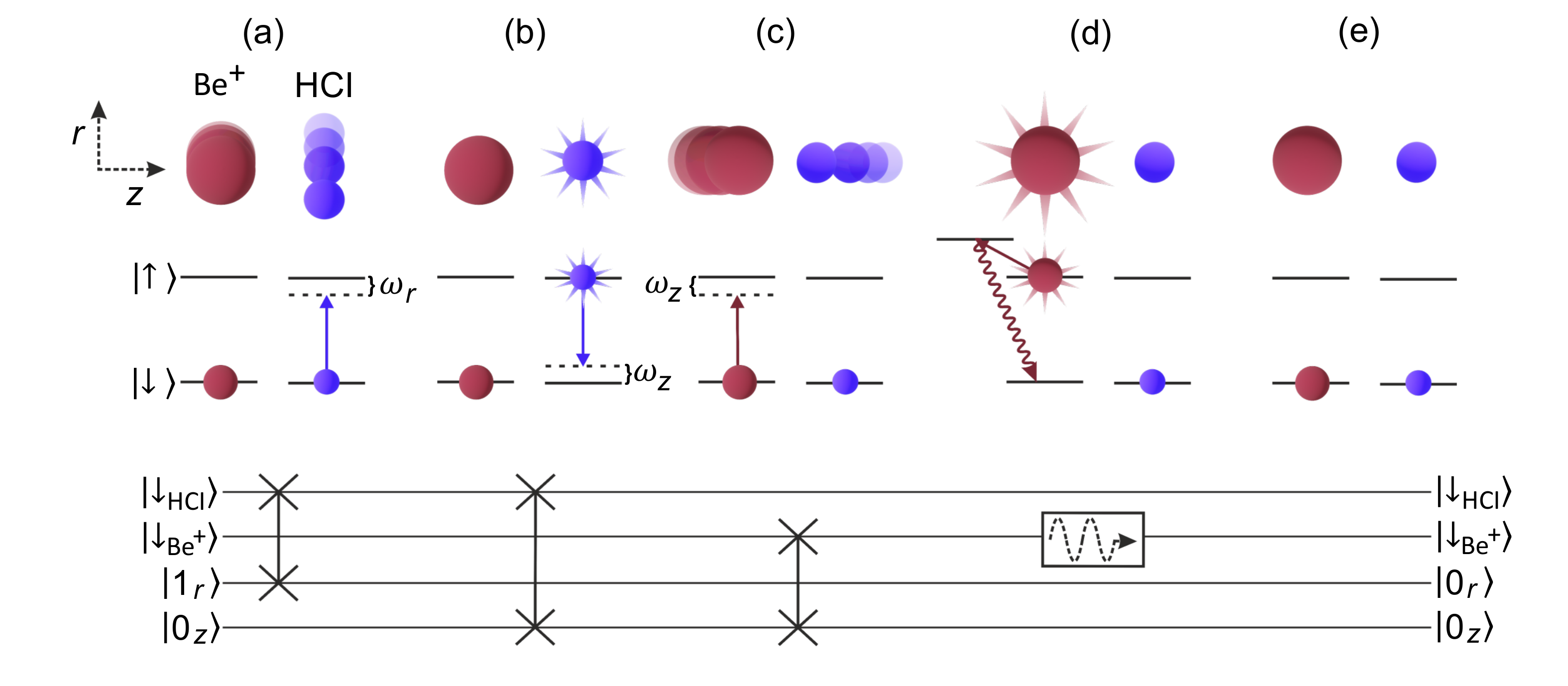}
	\caption{Upper: Scheme of the quantum logic sideband cooling sequence representing the state of the ion crystal before each of the depicted laser pulses is applied. Pulses (a) and (b) on the HCI (purple symbols) swap a phonon from one of the WCR modes ($r$) into the axial out-of-phase mode ($z$) through the excited electronic state of the HCI. Pulse (c) on the Be$^{+}$ ion (red symbols) removes the added phonon from the axial out-of-phase mode and converts it into an electronic excitation of the Be$^{+}$ ion. Pulse (d) dissipatively resets the Be$^{+}$ to its initial electronic state through spontaneous decay, completing the sequence and ensuring unidirectionality and cooling. Solid arrows depict driven transitions, undulating arrows represent spontaneous decay. $\ket{\downarrow}$ and $\ket{\uparrow}$ represent the ground and excited electronic states of the relevant ion, respectively. Lower: Representation of the sequence as a quantum circuit composed of a series of SWAP quantum gates. $\ket{1_{i}}$ and $\ket{0_{i}}$ represent the presence and lack of a phonon in mode $i$, respectively.}
	\label{fig:qlsbc}
\end{figure*}

Despite Be$^{+}$ having the highest charge-to-mass ratio of any singly-charged ion with a suitable laser-cooling transition, its mismatch to Ar$^{13+}$ is so great that the amplitudes of motion for the Be$^{+}$ ion in the radial in-phase modes are two orders of magnitude smaller than in the other four modes \cite{kozlov_hci, supplementary}. This slows cooling since the lasers addressing the Be$^{+}$ ion cannot efficiently remove energy from these weakly-coupled radial (WCR) modes. Furthermore, operations on the WCR modes by resolved sideband techniques suffer from the extremely weak coupling strength on the Be$^{+}$ ion. The decoupling is much less severe in the axial direction of the trap, where the confinement is provided by a static dc potential. Hence, this direction is generally preferred for quantum logic operations because they can be performed using either ion. \par

The laser beams for manipulation of the two-ion crystal are shown in Fig.~\ref{fig:beams}. They are delivered in the horizontal plane, and therefore each beam projects equally onto the $x$ and $y$ axes of the trap. We use a 441~nm laser for performing operations on the Ar$^{13+}$ by driving the $^2$P$_{1/2} \rightarrow {}^2$P$_{3/2}$ fine-structure transition. The laser can enter from two possible directions in order to couple to either radial or axial motion of the ion crystal as required. The Doppler cooling and repumping lasers have wavelengths near 313~nm for driving the $^2$S$_{1/2} \rightarrow {}^2$P$_{3/2}$ and $^2$S$_{1/2} \rightarrow {}^2$P$_{1/2}$ transitions in the Be$^{+}$ ion, respectively. They intersect the $z$ axis at an angle of $30^{\circ}$, and are precisely aligned with the quantization axis, defined by an applied magnetic field with a flux density of approximately 20~\micro{T} (200~mG).\par

Operations on the axial motional modes using the Be$^{+}$ ion employ stimulated Raman transitions between the $F=2$ and $F=1$ hyperfine sublevels of the ${}^2$S$_{1/2}$ state. We drive them with two beams derived from a third 313~nm laser; each beam intersects the trap $z$ axis at an angle of $30^{\circ}$, with the effective wavevector of the two beams lying along the $z$-axis.\par

\section{Experimental sequence}
\label{sec:sequence}

The main experimental sequence runs as follows. We first apply 200~ms of Doppler cooling to ensure effective cooling of the WCR modes (see section \ref{sec:recoil}). The two axial modes are then cooled close to their ground states using stimulated Raman transitions driven on the Be$^{+}$ ion. After that, we optically pump the HCI into the desired ground state using quantum-logic-assisted state preparation \cite{micke_coherent_2020}. The WCR modes are then optionally cooled using the quantum algorithm shown in Fig.~\ref{fig:qlsbc}, which relies on driving resolved sidebands of the transitions in each of the ions that are either lower in energy than the carrier (red-sideband transition, RSB) or higher in energy (blue-sideband transition, BSB). This is achieved by detuning the involved laser away from the carrier in the appropriate direction by the respective motional mode frequency. (a) The excitation of the WCR mode is coherently mapped onto the electronic state of the HCI by driving the ${}^2$P$_{1/2} \rightarrow {}^2$P$_{3/2}$ magnetic-dipole transition in the Ar$^{13+}$ ion using a laser beam with projection purely onto the radial trap axes. The laser is tuned to the RSB of the WCR mode to be cooled, and results in the removal of one phonon from this mode if the excitation is successful. (b) A second pulse from the same laser maps the electronic excitation of the Ar$^{13+}$ ion onto the strongly-coupled axial out-of-phase motional mode. It is implemented by applying a beam with a projection purely onto the axial direction, with the laser frequency tuned to the RSB of the axial out-of-phase motion. This adds a phonon to the axial out-of-phase mode only if pulse (a) had removed a phonon from the WCR mode. (c) This axial motional excitation is then mapped onto an electronic excitation in the Be$^{+}$ ion by driving the appropriate RSB of the stimulated Raman transition on the Be$^{+}$ ion. (d) Dissipation and irreversibility as required for cooling is provided by resetting the Be$^{+}$ ion to its initial electronic state through spontaneous decay from the ${}^{2}$P$_{1/2}$ level after excitation by the repumping laser, completing the cycle. We then repeat steps (a)-(d) for the second WCR mode, and this whole sequence several times until (e) a temperature close to the motional ground state is reached. In practice, we cycle through steps (c) and (d) several times to neutralize the recoil of the ion crystal due to spontaneous photon scattering \cite{wan_efficient_2015, chen_sympathetic_2017, che_efficient_2017, leopold_cryogenic_2019}, and a short optical pumping step for the HCI is implemented before each cycle to ensure that the correct electronic state has been prepared beforehand. An advantage of the weak coupling is that, unlike the other four modes of the crystal, the WCR modes suffer very little heating from spontaneous decay of the Be$^{+}$ ion, as we elaborate upon in section \ref{sec:recoil}.\par
The first two pulses (a) and (b) on the Ar$^{13+}$ ion could also in principle be joined to implement a direct SWAP gate between motional excitation of the WCR mode and the axial mode by employing a Raman coupling. This would be advantageous in systems with excited electronic state lifetimes shorter than the gate time, or where population of the excited state needs to be avoided.\par

\begin{figure}
	\includegraphics[width=3.4in]{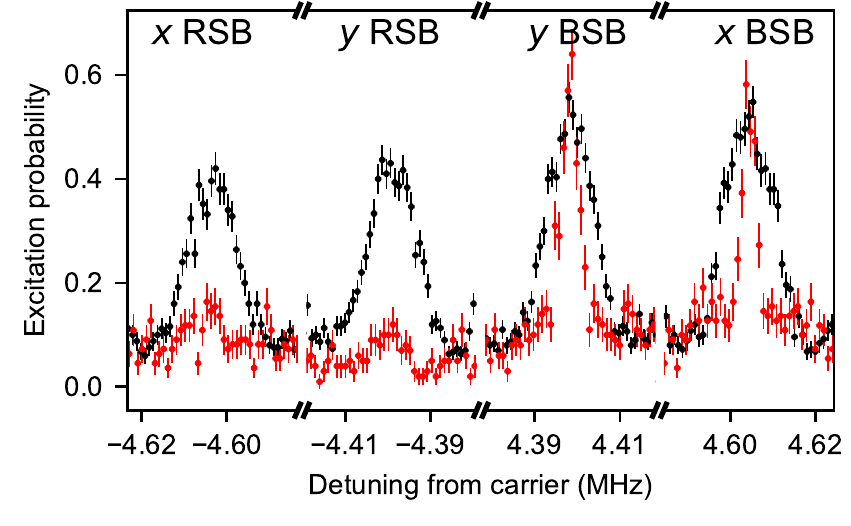}
	\caption{WCR motional sidebands of the two-ion crystal observed on the HCI before (black) and after (red) resolved sideband cooling. The error bars are statistical only, governed by quantum projection noise. The observed difference in the full-width-at-half-maximum values between the two cases is caused by the different lengths of the interrogation pulse used, which were adjusted in each case to produce the highest possible contrast. The scans have been re-centered for presentation, correcting for small drifts in the secular frequencies.}
	\label{fig:sidebands}
\end{figure}

Fig.~\ref{fig:sidebands} shows a sideband spectrum measured on the HCI after eight cooling pulses on each WCR mode, shown together with the Doppler-cooled case for comparison. The final state of the HCI after the interrogation pulse is determined using quantum logic \cite{schmidt_spectroscopy_2005}, which resembles a simplified version of the algorithmic cooling sequence using only pulses (b) and (c) to map the electronic excitation of the HCI onto the electronic state of the Be$^{+}$, followed by detection of the electronic state of the Be$^{+}$ using state-dependent fluorescence \cite{micke_coherent_2020}. The mean phonon occupation number $\langle n \rangle$ can be obtained from the asymmetry between the peak RSB and BSB excitation probabilities \cite{monroe_resolved-sideband_1995}, with a small correction applied owing to the finite background signal caused by imperfect ground-state cooling of the axial mode used for the quantum logic operation. The slight asymmetry after Doppler cooling implies mean occupation numbers of 3.1(13) and 3.3(13) phonons in the $x$ and $y$ modes respectively, where the number in brackets represents the $1\sigma$ uncertainty on the last significant figure. These values are close to the calculated value \cite{supplementary} of $\langle n \rangle \approx3$. The strong suppression of the red sidebands after ground-state cooling indicates preparation of a nearly pure motional ground state, with mean occupation numbers in the $x$ and $y$ modes of 0.39(15) and 0.21(8) phonons. If a thermal distribution of states is assumed \cite{wineland_experimental_1998}, this is equivalent to probabilities of approximately 72\% and 83\% of occupying the ground states of the two modes, or residual temperatures of $T\approx174$~\micro{K} and 121~\micro{K} above the zero-point energy. These values are close to what would be expected from our experimental parameters, and could be further improved with modifications to the cooling sequence \cite{supplementary}.

\section{Demonstration of the weak coupling}
\label{sec:recoil}

\begin{figure}
	\includegraphics[width=3.4in]{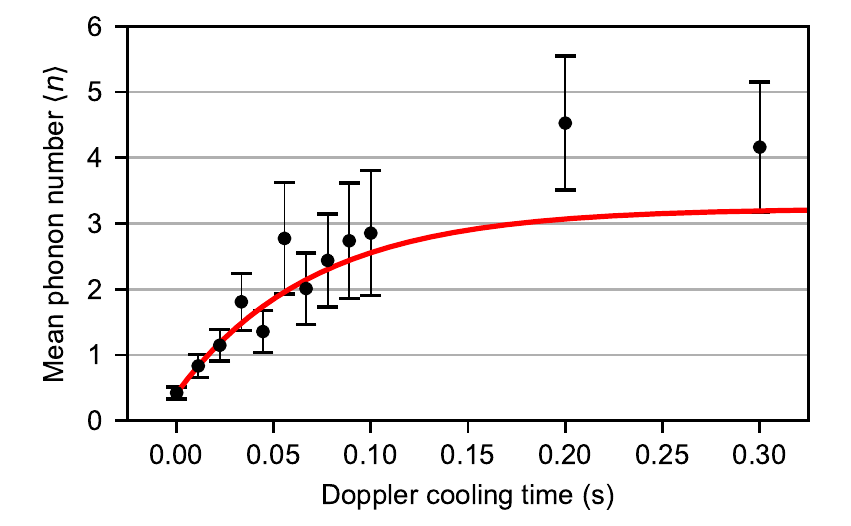}
	\caption{Heating of the $y$ WCR mode out of the ground state during the Doppler cooling pulse applied to the Be$^{+}$ ion (black points), with the expected background from anomalous heating removed. The data is fitted (red line) using a semiclassical model.}
	\label{fig:doppler_heating}
\end{figure}

We experimentally demonstrate the weak coupling of the Be$^{+}$ cooling laser to the WCR modes by measuring the time constant for Doppler cooling of these modes. Precise measurements of values of $\langle n \rangle \gtrsim 5$ are difficult, however, owing to the very small asymmetry between the red and blue sidebands at these temperatures. We therefore reverse the process and measure heating of the ion crystal out of the ground state by the Doppler cooling laser. We first apply a 200~ms-long Doppler cooling pulse to ensure that the ion crystal is close to the expected equilibrium temperature. Both WCR modes are then prepared close to their ground states, as described in section \ref{sec:experiment}. As the WCR modes are now below the equilibrium temperature for Doppler cooling, applying the Doppler cooling and repump lasers at this point will add heat to the crystal and drive it back towards the equilibrium temperature. Varying the length of this pulse allows the time constant for the heating to be measured. The cooling laser is red-detuned from resonance by the half-width of the transition linewidth and set to approximately 0.3 times the saturation intensity. After the heating pulse, the axial modes are cooled to the ground state using the Be$^{+}$ ion, and the amplitude of either the red or blue sideband of the WCR mode under test is then measured as described in section \ref{sec:experiment}. We interleave the measurements of the red and blue sidebands on a cycle-by-cycle basis to reduce sensitivity to drifts. Mode temperatures reach a steady-state value for long Doppler cooling pulses. We fit the data by means of a semiclassical model \cite{stenholm_semiclassical_1986, eschner_laser_2003} with our approximate experimental parameters, but with the Lamb-Dicke parameter and initial value for $\langle n \rangle$ as fit parameters, with the results shown in Fig.~\ref{fig:doppler_heating}. The fitted Lamb-Dicke parameter is $3.7(4)\times10^{-3} \cdot \cos{(\theta_{\textrm{cool}})}$, where $\theta_{\textrm{cool}} \approx 69^{\circ}$ is the intersection angle between the cooling laser and the radial modes \cite{supplementary}. The quoted uncertainty is only from the fit and does not include additional uncertainties in experimental parameters. This is in good agreement with the calculated value \cite{supplementary} of $3.3\times10^{-3} \cdot \cos{(\theta_{\textrm{cool}})}$. The $1/e$ time constant for the heating is 70~ms, and equilibrium is only reached for cooling times in the range of hundreds of milliseconds.\par

\section{Limitations to the cooling process}
\label{sec:limitations}

The main limitation to the rate at which energy can be removed by the algorithmic cooling scheme is the duty cycle resulting from dead time and overhead such as axial ground state cooling and optical pumping steps among others. Each step is limited by the shortest possible pulse durations for each part of the process, and the required number of repetitions of each step. The minimum pulse length for both ions is limited by off-resonant coupling to other nearby transitions that are not part of the cooling process, such as the much stronger carrier transitions that change only the electronic state \cite{monroe_resolved-sideband_1995}. Temporal shaping of the laser pulse can mitigate this to some degree by limiting the high-frequency content of the laser spectrum \cite{harris_use_1978}. Ultimately, when the Rabi frequency approaches the splitting between the lines, the resulting line broadening will cause the features to merge. In our experiment we employ laser pulse durations on the order of 20~\micro{s} for the Raman operations for the Be$^{+}$ ion, and 180~\micro{s} for operations on the HCI, resulting in respective full-width-at-half-maximum (FWHM) values of 40~kHz and 4.4~kHz for the observed motional sidebands on the two ions. The FWHM values for the carrier transitions are approximately one order of magnitude larger in both cases, but they are still well-resolved from the motional sidebands.\par

The number of repetitions of each step of the algorithmic cooling process depends on experimental imperfections in implementing the gates, including inefficiencies in optical pumping. Additional time is also needed for the frequent reprogramming of frequency generators used to tune the laser frequencies by means of acousto-optic modulators. In our current implementation, a single algorithmic cooling cycle has a duration of 4.5~ms. If both WCR modes are cooled in parallel by interleaving cycles that individually address the $x$ and $y$ modes, a maximum cooling rate $\Gamma_{c}$ of approximately 111 phonons per second could be achieved for both modes. This assumes that the appropriate electronic state of the HCI and the ground states of the axial modes of the crystal have been prepared in advance. Despite the significant dead time in our algorithmic cooling cycle, the achievable cooling rate is still an order of magnitude higher than the Doppler cooling rate near to equilibrium (see section~\ref{sec:recoil}), and allows much lower temperatures to be reached.\par

A key parameter governing the equilibrium temperature after ground state cooling of the WCR modes is the anomalous heating rate $\Gamma_{h}$ caused by fluctuating electric fields at the position of the ions. The heating rates of the ($x, y$) modes of a single Be$^{+}$ ion in this trap were previously measured to be (1.9(3), 0.7(2)) phonons per second at mode frequencies of (2.5, 2.2)~MHz \cite{leopold_cryogenic_2019}. Significantly higher heating rates will be observed for modes where the HCI motion is dominant, however, since the high charge state of the HCI leads to a much stronger coupling to electric field noise (for a mode dominated by a single ion with charge $Z$, $\Gamma_{h} \propto Z^{2}$ \cite{wineland_experimental_1998-1, brownnutt_ion-trap_2015}).\par

\begin{table}
	\begin{center}
		\begin{tabular*}{3.375in}{c @{\extracolsep{\fill}} ccc}
			
			Mode & $\omega/2\pi$  & $\Gamma_{h}$  & $S_{E}(\omega)$  \\
			& (MHz) & (quanta/s) & (V$^{2}$m$^{-2}$Hz$^{-1}$)\\
			\hline
			WCR $x$ & 4.59 & 9.9(10) & $1.8(2)\times10^{-15}$\\ 
			WCR $y$ & 4.41 & 2.0(2) & $3.5(3)\times10^{-16}$\\ 
			
		\end{tabular*}
		\caption{\label{tab:heating} Properties of the WCR modes of the two-ion crystal: frequency $\omega/2\pi$, heating rate $\Gamma_{h}$, and calculated electric field noise power spectral density at this frequency $S_{E}(\omega)$.}
	\end{center}
\end{table}

For measuring the anomalous heating rates of the WCR modes, we first cool them both to the motional ground state as described in section \ref{sec:sequence}. They are then allowed to heat freely for periods of up to 0.5~seconds. After the chosen delay, we determine the mean number of phonons per mode by measuring the excitation probabilities on the red and blue sidebands using quantum logic \cite{monroe_resolved-sideband_1995}.
The heating rates, summarized in Table \ref{tab:heating}, can be converted to a single-sided power spectral density for the electric-field noise that would cause an equivalent heating, assuming that it arises from noise at $\omega$ and not from coupling to micromotion sidebands at frequencies $\Omega$ and $\Omega\pm\omega$ \cite{brownnutt_ion-trap_2015}. The value of $3.5(3)\times10^{-16}$~V$^{2}$m$^{-2}$Hz$^{-1}$ measured for the WCR $y$ mode is to our knowledge the lowest value ever reported for a radiofrequency ion trap \cite{niedermayr_cryogenic_2014, brownnutt_ion-trap_2015}. The reduction in noise compared to the older single-ion data given earlier is likely due to subsequent improvements in the rf and dc circuitry used to drive the trap. For comparison against lowly-charged systems, this level of noise would lead to a heating rate of 0.012 quanta/s for a single $^{40}$Ca$^{+}$ ion at this motional frequency. Identifying the origin for the difference in noise spectral density for the two radial directions will be subject of future work.

\section{Keeping the HCI in the ground state}
\label{sec:constantcooling}
\begin{figure}[t!]
	\includegraphics[width=3.4in]{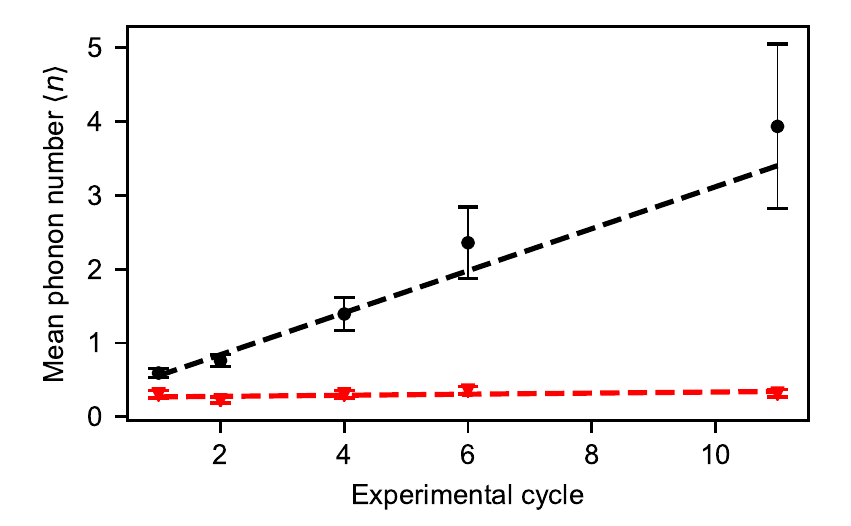}
	\caption{Suppression of the heating of the $x$ WCR mode. Without additional algorithmic cooling, the ion heats up due to the anomalous electric field noise (black circles). A linear fit (black dashed line) implies a heating rate of 8.9(16) phonons/second, consistent with the previously measured value for this mode (see Table \ref{tab:heating}). With one algorithmic cooling cycle applied on each WCR mode per experimental cycle, this heating is suppressed (red triangles), with a fitted heating rate of 0.18(15) phonons/second (red dashed line).}
	\label{fig:steady_state_gsc}
\end{figure}

Under normal operation, it is desirable to minimize dead time by keeping the Doppler cooling pulses significantly shorter than the 200~ms value used for the measurements above. The disadvantage of this is that anomalous heating of the WCR modes is not suppressed by the short cooling pulses. Therefore, the WCR modes gradually heat up until the increased Doppler cooling rate caused by the higher ion temperature can balance the anomalous heating \cite{stenholm_semiclassical_1986, eschner_laser_2003}. We demonstrate this by preparing the two modes in the ground state and then performing a typical `clock' experiment. The clock cycle resembles the experiment detailed in section \ref{sec:recoil}, but with two main modifications: (1) omitting the additional pulse from the cooling laser that was used to cause heating out of the ground state, and (2) rather than measuring the red and blue sidebands after a single cycle, we perform a variable number of cycles (each of duration 32~ms) before reading out the ion temperature. Each cycle contains a total Doppler cooling time of less than 1~ms. Without algorithmic cooling, it can be seen that the mode temperature slowly increases between cycles in line with the expected anomalous heating rate. This is shown in Fig.~\ref{fig:steady_state_gsc}, where the temperature of the $x$ WCR mode is displayed as a `worst-case' (since we observe a higher heating rate for this mode than for the $y$ WCR mode). The anomalous heating rates of only a few quanta per second can be suppressed by only occasionally applying red sideband pulses: we add a single algorithmic cooling cycle for each of the two WCR modes to the end of our normal optical pumping routine for the HCI \cite{micke_coherent_2020} while repeating the above measurement. As discussed in section \ref{sec:limitations}, this adds only 9~ms of overhead to the experimental cycle, but nevertheless the effect of anomalous heating of the WCR modes can be completely suppressed as the heating rates are lower than the maximum cooling rate of 24 phonons per second under these conditions.\par

\section{Conclusions}
To conclude, we have achieved algorithmic ground-state cooling of the radial in-phase motional modes of a two-ion Coulomb crystal containing a HCI, despite the exceptionally weak coupling between the ions in these modes. In conjunction with the ground-state cooling of the axial modes of the crystal, this is the coldest HCI prepared in a laboratory thus far. The technique demonstrated here is very general, and could be applied to a plethora of ions that cannot be directly laser-cooled and would have an unavoidably large charge-to-mass ratio mismatch with their cooling ion, as is the case for (anti-)protons \cite{smorra_base_2015, Bohman2018}, highly charged ions \cite{kozlov_hci, schmoger_coulomb_2015, micke_coherent_2020} and trapped charged macroscopic particles, such as nanospheres \cite{millen_cavity_2015, delord_diamonds_2017, bykov_laser_2019}, graphene \cite{kane_levitated_2010, nagornykh_cooling_2015} or nanodiamonds \cite{kuhlicke_nitrogen_2014, conangla_motion_2018}.\par
For the discussed Be$^{+}$-Ar$^{13+}$ system, if the mean phonon numbers in the WCR modes could be kept below the conservative target of $\langle n_{x,y} \rangle = 0.5$, time dilation from the residual ion velocity in these modes would lead to a total fractional systematic shift of only $-1 \times 10^{-18}$ on the Ar$^{13+}$ transition resonance frequency. This eliminates the final obstacle for the development of an optical frequency standard based on highly charged ions \cite{kozlov_hci} with an accuracy that could surpass that of the best optical frequency standards available today \cite{mcgrew_atomic_2018, sanner_optical_2019, brewer_27+_2019, bothwell_jila_2019}.\par


%
%

%


\vspace{2 mm}
\acknowledgments{The authors would like to thank Erik Benkler and Thomas Legero for their contributions to the frequency stabilization of the HCI spectroscopy laser, Giorgio Zarantonello for fruitful discussions, and Ludwig Krinner for helpful comments on the manuscript. The project was supported by the Physikalisch-Technische Bundesanstalt, the Max-Planck Society, the Max-Planck–Riken–PTB–Center for Time, Constants and Fundamental Symmetries, and the Deutsche Forschungsgemeinschaft (DFG, German Research Foundation) through SCHM2678/5-1, the collaborative research centres SFB 1225 ISOQUANT and SFB 1227 DQ-mat, and under Germany’s Excellence Strategy – EXC-2123 QuantumFrontiers – 390837967. This project 17FUN07 CC4C has received funding from the EMPIR programme co-financed by the Participating States and from the European Union’s Horizon 2020 research and innovation programme. S.A.K. acknowledges financial support from the Alexander von Humboldt Foundation.}


%

\end{document}



\title{Supplemental Material for Algorithmic Ground-state Cooling of Weakly-Coupled Oscillators using Quantum Logic} 



\author{Steven A. King}
\email[Contact: ]{steven.king@ptb.de}
\affiliation{Physikalisch-Technische Bundesanstalt, Bundesallee 100, 38116 Braunschweig, Germany}

\author{Lukas J. Spieß}
\affiliation{Physikalisch-Technische Bundesanstalt, Bundesallee 100, 38116 Braunschweig, Germany}

\author{Peter Micke}
\affiliation{Physikalisch-Technische Bundesanstalt, Bundesallee 100, 38116 Braunschweig, Germany}
\affiliation{Max-Planck-Institut f\"ur Kernphysik, Saupfercheckweg 1, 69117 Heidelberg, Germany}

\author{Alexander Wilzewski}
\affiliation{Physikalisch-Technische Bundesanstalt, Bundesallee 100, 38116 Braunschweig, Germany}

\author{Tobias Leopold}
\affiliation{Physikalisch-Technische Bundesanstalt, Bundesallee 100, 38116 Braunschweig, Germany}

\author{Jos\'e~R.~{Crespo~L\'opez-Urrutia}}
\affiliation{Max-Planck-Institut f\"ur Kernphysik, Saupfercheckweg 1, 69117 Heidelberg, Germany}

\author{Piet O. Schmidt}
\affiliation{Physikalisch-Technische Bundesanstalt, Bundesallee 100, 38116 Braunschweig, Germany}
\affiliation{Institut f\"ur Quantenoptik, Leibniz Universit\"at Hannover, Welfengarten 1, 30167 Hannover, Germany}

\date{\today}

\pacs{}

\maketitle 


In this supplemental material, we describe in more detail the experimental setup and elaborate upon some of the techniques and calculations used in the main paper.

\section{Experimental setup}
This section is dedicated to the coherent laser manipulation of Be$^+$ and the major changes to the apparatus since publication of Ref.~\cite{leopold_cryogenic_2019}.

\label{sec:experiment}

\begin{figure}
	\includegraphics[width=3.4in]{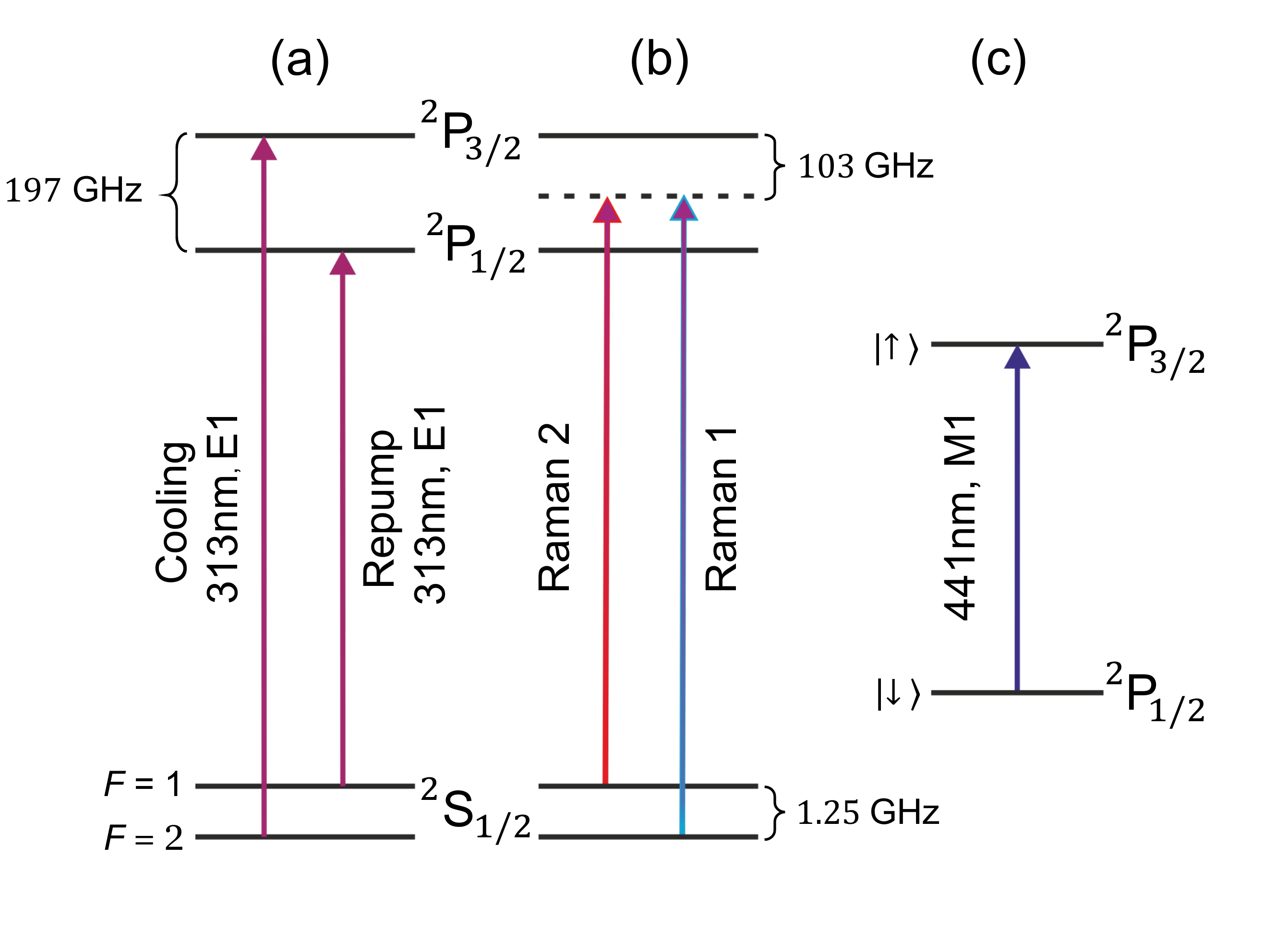}
	\caption{Partial term schemes of $^{9}$Be$^{+}$ and $^{40}$Ar$^{13+}$ (energies not to scale), showing the transitions used for manipulation of the two-ion crystal. (a) Doppler cooling and repumping using the Be$^{+}$ ion, (b) driving stimulated Raman transitions on the Be$^{+}$ ion, and (c) driving sideband transitions using the Ar$^{13+}$ ion. Hyperfine sublevels of the $^{2}$P$_{1/2}$ and $^{2}$P$_{3/2}$ states of $^{9}$Be$^{+}$ are omitted for clarity.}
	\label{fig:term_schemes}
\end{figure}

\subsection{Laser systems for Be$^{+}$}
The relevant level structures of the Be$^{+}$ and Ar$^{13+}$ ions are shown in Fig.~\ref{fig:term_schemes}. Sideband operations on the Be$^{+}$ ion are performed using stimulated Raman transitions driven between the $F=2$~$(m_{F}=2)$ and $F'=1$~$(m_{F'}=1)$ sublevels of the $^2$S$_{1/2}$ state. The Raman beams are derived from the same 313~nm source, which has a detuning of approximately $-103$~GHz from the $^2$S$_{1/2} \rightarrow {}^2$P$_{3/2}$ transition. Operating at this detuning allows cancellation of the induced ac Stark shifts from the individual beams to a high degree, as they have a roughly equal and opposite detuning from the $^2$S$_{1/2} \rightarrow {}^2$P$_{1/2}$ transition. However, this leads to a reduction of the Rabi frequencies caused by destructive interference between the two pathways \cite{wineland_quantum_2003, ozeri_hyperfine_2005}. The beams have a relative detuning near 1.25~GHz, which matches the hyperfine splitting in the $^2$S$_{1/2}$ ground state. Variation of this parameter allows the addressing of other spectral features. The first Raman beam (`Raman 1' in Fig.~\ref{fig:term_schemes}) and the second Raman beam (`Raman 2') respectively address the $F=2$ and  $F=1$ hyperfine sublevels. The second Raman beam counter-propagates with respect to the cooling laser. It has $\sigma^{+}/\sigma^{-}$ polarization in order to minimize the ac Stark shift induced by this beam \cite{wineland_quantum_2003}. The first Raman beam can be delivered from one of two possible directions, ensuring that in combination with the second Raman beam the effective projection of the two beams is either axial or radial, as required. Due to the geometry of the vacuum chamber, the two possible directions for the first Raman beam are also both oriented at an angle of 30$^{\circ}$ to $z$, such that that they can not have pure $\pi$-polarization. Their associated Stark shifts can still be reduced to the kHz-level by careful tuning of the laser polarization, which is only a few percent of the 40~kHz transition linewidths produced under our typical operating conditions.\par

After the driving of a stimulated Raman transition, an independent 313~nm laser source \cite{king_self-injection_2018} is used to recycle (repump) the Be$^{+}$ ion to the ${}^{2}$S$_{1/2}, F=2$ state by excitation to the ${}^{2}$P$_{1/2}$ level \cite{leopold_cryogenic_2019, micke_coherent_2020} from which it can spontaneously decay to the desired state.\par

\subsection{Stabilization and monitoring of laser powers}
Relatively large shifts in laser frequency of up to 10~MHz are required to probe the motional sidebands of the atomic transitions. This leads to changes in the diffraction efficiency of the acousto-optic modulators (AOM) used to tune the lasers. To ensure identical laser powers and hence stabilize the Rabi frequencies and laser-induced systematic shifts of the resonances, we stabilize them on a pulse-by-pulse basis using signals from photodiodes located near to the ion trap. Feedback is then applied to the drive power of the appropriate AOM using a STEMlab (formerly Red Pitaya)-based system \cite{hannig_highly_2018} in a sample-and-hold configuration. The status of these sample-and-hold systems along with various other systems such as laser locks are monitored on a cycle-by-cycle basis by our experiment-control system. Cycles where one or more parameters were outside tolerable ranges are discarded and repeated.\par

\subsection{Temporal shaping of laser pulses}
As a consequence of the small applied magnetic field (20~\micro{T}) and resulting first-order Zeeman shifts, impure laser polarizations, low motional frequencies ($\sim 1$~MHz) and high Lamb-Dicke parameters in the axial direction, the spectra of both the HCI and Be$^{+}$ ion observed in this direction have many closely separated lines including pronounced intermodulation peaks. To reduce the probability of off-resonant driving of unwanted nearby transitions whilst maintaining high Rabi frequencies, almost all laser pulses coupling to first-order motional sidebands in this direction are temporally shaped such that the Rabi frequency evolves with a profile that closely resembles a Blackman waveform, thereby suppressing the problematic `wings' of the Rabi line profile \cite{harris_use_1978}.\par

\subsection{Stabilization of rf trapping voltage}
As the trap depth must be reduced in order to reload a new HCI after a charge-exchange collision \cite{micke_closed-cycle_2019, micke_coherent_2020}, the radial motional frequencies are prone to drift after returning to the typical values used during the presented experiments. This is likely attributable to the increased rf power dissipation causing changes in the temperature and hence the electrical conductivity of both the trap and rf resonator which are mounted within the cryostat \cite{leopold_cryogenic_2019}. To avoid this, the rf trap depth is actively stabilized by monitoring the pickup on an antenna next to the ion trap used for manipulation of the Be$^{+}$ ion using microwave fields. Feedback is applied to the amplitude of the rf synthesizer used to drive the trap in an approach similar to that of reference \cite{johnson_active_2016}. The resulting radial secular frequencies are typically stable at the fractional level of $10^{-3}$.

\section{Mode temperatures after Doppler cooling}
\label{sec:temperature}

The cooling laser is delivered in the horizontal plane at an angle of 30$^{\circ}$ to the trap $z$ axis, and therefore has a significantly weaker projection onto the radial modes of the Coulomb crystal than onto the axial modes, with an intersection angle of $\theta_{\textrm{cool}} = \arccos(\sin30^{\circ} \cdot \cos45^{\circ}) \approx 69^{\circ}$. This leads to an increase in the equilibrium temperature of the radial modes after Doppler cooling compared to the ideal case. Using the notation of reference \cite{itano_laser_1982}, in the absence of anomalous heating, the total energy after Doppler cooling for mode $i$ can be expressed as:

\begin{equation}
\left<E_{i}\right> = 2\langle E_{\mathrm{kin},i} \rangle = \left( 1 + \frac{f_{si}}{f_{i}} \right) \frac{\hbar \Gamma}{4},
\end{equation}

where $\langle E_{\mathrm{kin},i} \rangle$ is the kinetic energy in mode $i$, $f_{si}$ is a factor determined by the angular distribution of  the spontaneously emitted photons, $f_{i}$ is a factor determined by the projection of the cooling laser onto the direction of motion, $\hbar$ is the reduced Planck constant, and $\Gamma$ is the linewidth of the transition used for laser cooling. This can be equated with the energy of the quantum harmonic oscillator in terms of the mean number of phonons $\langle n_{i} \rangle$:

\begin{equation}
\label{eq:nbar}
\langle E_{i} \rangle = \left(\langle n_{i} \rangle +\frac{1}{2}\right) \hbar \omega_{i}.
\end{equation}

For typical frequencies of $\omega = 2\pi \times 4.5$~MHz for the WCR modes in this experiment, the non-ideal orientation of the Doppler cooling beam ($f_{x,y} \approx 0.35$), non-isotropic emission due to only $\sigma^{+}$-polarized photons being scattered during Doppler cooling ($f_{sx,sy} \approx 0.31$), and $\Gamma=2\pi\times18$~MHz linewidth for the cooling transition in Be$^{+}$ lead to a mean vibrational quantum number of $\langle n_{x,y} \rangle \approx3$ after Doppler cooling, approximately a factor of two higher than the expected value of $\langle n_{x,y} \rangle \approx 1.5$ under idealized conditions where the cooling beam has an equal projection onto all axes and the spontaneous emission is isotropic.\

\section{Normal mode amplitudes}
\label{sec:modes}

\begin{table}
\begin{center}
\begin{tabular*}{3.375in}{c @{\extracolsep{\fill}} cccc}

 Axis & Mode & $\omega/2\pi$  & Be$^{+}$ amplitude & Ar$^{13+}$ amplitude  \\
 & & (MHz) & (nm) & (nm)\\
 \hline
 $x$ & IP (WCR) & 4.62 & 0.15 & 5.2\\
 $x$ & OP (SCR) & 1.39 & 20 & 0.13\\
 $y$ & IP (WCR) & 4.42 & 0.17 & 5.4\\
 $y$ & OP (SCR) & 1.15 & 22 & 0.15\\ 
 $z$ & IP & 1.15 & 18 & 6.3\\ 
 $z$ & OP & 1.56 & 11 & 7.2\\ 
\end{tabular*}
\caption{\label{tab:modeamplitudes} Calculated frequencies of the axial ($z$) and radial ($x, y$) normal modes for the two ions in a Be$^{+}$ - Ar$^{13+}$ Coulomb crystal under our approximate experimental trapping conditions, along with the extents of the ground-state wavefunctions for the individual ions in these modes. IP and OP refer to in- and out-of-phase motion, respectively, and SCR and WCR refer to the strongly-coupled and weakly-coupled radial modes, respectively. The Be$^+$ ion has a much smaller motional amplitude in the $x$ and $y$ IP modes, leading to greatly inhibited laser cooling of these modes.}
\end{center}
\end{table}

In this section, we present the calculated extents of the ground-state wavefunctions for the two ions in each of the six normal modes of the Coulomb crystal, including the amplitude scaling factors arising from the unequal distribution of energy in the modes between the ions. These were calculated for our typical experimental conditions by extending the approaches of references \cite{james_quantum_1998, kielpinski_sympathetic_2000, wubbena_sympathetic_2012} to include the different charges of the two ions. The Lagrangian for the system (including the trapping potential and Coulomb interaction) is solved to yield the normal modes of the ion crystal, in the usual coordinate system where the motion of the highly charged ion (HCI) is scaled by a factor $(m_{\textrm{Ar}}/m_{\textrm{Be}})^{-1/2}$. The amplitudes of motion $z_{i}$ for the two ions (with  $i \in \textrm{Be}^{+}, \textrm{Ar}^{13+}$) can be determined by treating them as quantum harmonic oscillators in the ground state of a given mode, yielding the relation:
\begin{equation}
z_{i} = |b_{\textrm{i}}| \sqrt{\frac{\hbar}{2 m_{\textrm{i}} \omega}},
\end{equation}
where $b_{i}$ is the appropriate component of the normalized eigenvector for the mode for ion $i$, $\hbar$ is the reduced Planck constant, $m_{i}$ is the ion mass, and $\omega / 2\pi$ is the secular frequency of the mode.\par
The results are summarized in Table \ref{tab:modeamplitudes}. The amplitude of only 0.15~nm for the Be$^{+}$ ion in the $x$ in-phase (IP) mode, referred to as the $x$ weakly-coupled radial (WCR) mode, corresponds in our setup to a Lamb-Dicke parameter of only $3.1\times10^{-3} \cdot \cos{(\theta_{\textrm{cool}})}$, more than two orders of magnitude smaller than the value of $0.4 \cdot \cos{(\theta_{\textrm{cool}})}$ for the $x$ out-of-phase (OP) mode. In the axial ($z$) direction, the ions do not display the same level of decoupling, and the laser cooling efficiency using the Be$^{+}$ remains high \cite{wubbena_sympathetic_2012}.

\section{Determining $\langle n \rangle$ from Motional Sideband Spectra}
\label{sec:nbar}
We derive the ground state population and heating rates from the mean phonon number $\langle n_{i} \rangle$ for mode $i$, which has been determined using the sideband ratio technique \cite{monroe_resolved-sideband_1995}. The value of $\langle n \rangle$ can be calculated from the observed contrasts $A$ on the red (RSB) and blue sidebands (BSB):

\begin{equation}
    \langle n \rangle = \frac{A_{\textrm{RSB}}/A_{\textrm{BSB}}}{1-A_{\textrm{RSB}}/A_{\textrm{BSB}}}.
\end{equation}
This ratio is independent of the employed Rabi probe time, as long as it is the same for RSB and BSB.

\subsection{Quantum logic background correction}
\label{subsec:qlbgd}
A finite background appears on the quantum logic signal even when the laser interrogating the HCI is far off-resonant or physically blocked. This arises from imperfect ground-state cooling of the axial out-of-phase mode, leading to a weak but non-zero red sideband on the Be$^{+}$ Raman transition. This is taken into account in all measurements, where a systematic correction of -1.0(5)\% is applied to all measured excitation probabilities to account for its value and typical instability. This correction is applied after the averaging of multiple datasets to avoid inadvertent and invalid averaging of this correction. The value of the background should be identical for the blue and red sideband measurements, as they are performed in an interleaved fashion. The uncertainty in the extracted value of $\langle n \rangle$ is therefore likely slightly overestimated as there is correlation in the background correction that is not accounted for when propagating the uncertainties.

\subsection{Scanning over the motional sidebands}
For Fig. 3 in the main text, the frequency of the laser addressing the HCI was scanned over the red and blue sidebands of the two WCR modes, and the line profiles were fitted. In order to minimize sensitivity to drifts in the motional frequencies, each red sideband and its blue counterpart were scanned in a narrow range around each feature. The individual probe frequencies were cycled through in a pseudo-random sequence, further reducing the correlation between successive measurements.\par

For the Doppler cooled-case, a simple squared-sinc function was used to fit the data described by a thermal superposition of lines with $n$-dependent Rabi frequencies. For the ground-state-cooled case, a Rabi line profile assuming a pure motional quantum state was fitted to the data. Line profile distortions, most noticeable in the wings of the excitation profile, arise from instability ($\sim 10^{-3}$) of the motional mode frequencies on the timescale of the experiments, dephasing from the residual population in higher Fock states, and noise on the excitation laser. To avoid the resulting underestimation of the fitted excitation amplitude and to be consistent with all other $\langle n\rangle$ measurements in this manuscript, we chose to use the offset-corrected (see subsection \ref{subsec:qlbgd}) experimental data point closest to the fitted line center for the calculation of $\langle n \rangle$, at the expense of increased statistical uncertainty.\par

\begin{figure*}[ht!]
	\includegraphics[width=\textwidth]{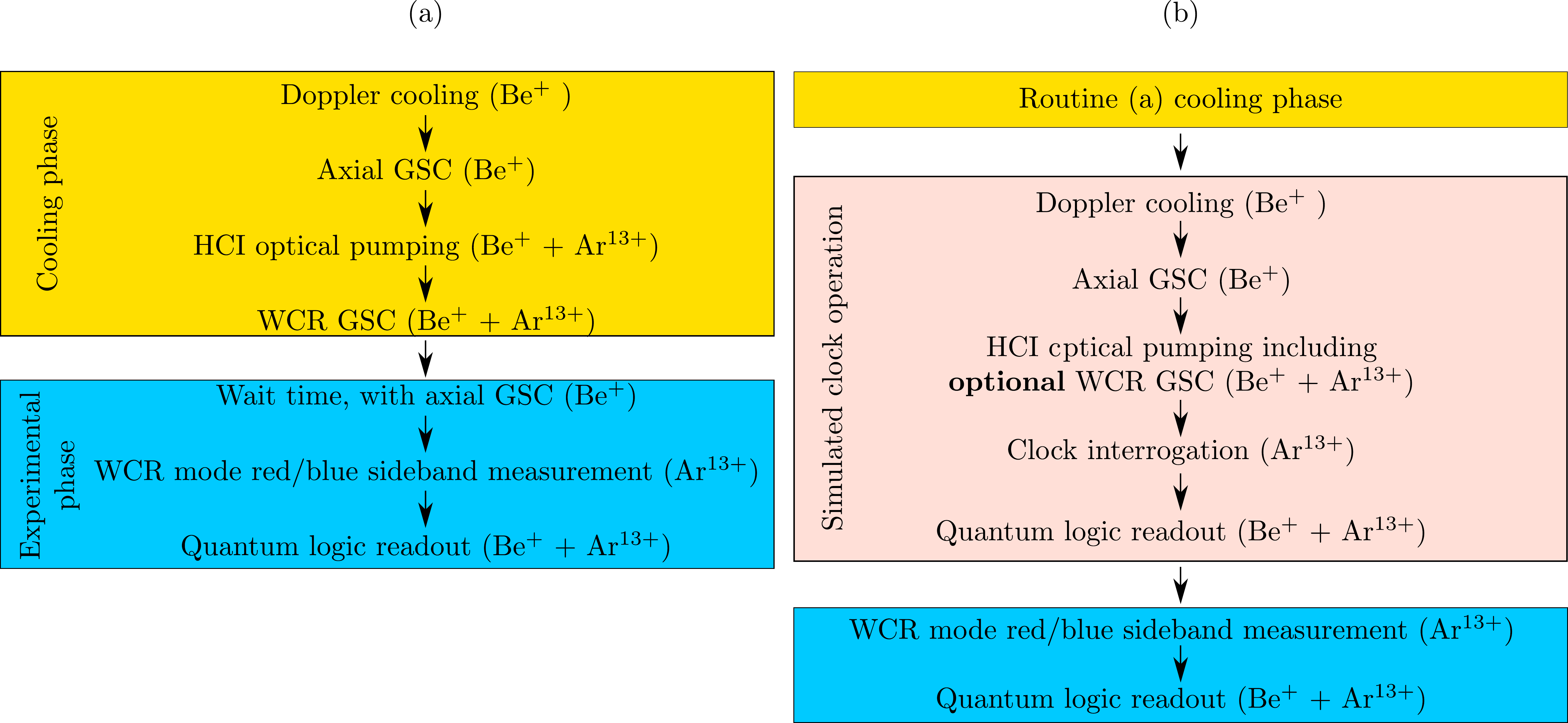}
	\caption{Simplified depiction of the two main experimental sequences used in this work, with the cooling phases (yellow), clock operation (pink) and experimental (blue) phases indicated.  GSC = ground-state cooling. The ions involved in each step are also indicated. Sequence (a) was used to measure anomalous heating rates for the WCR modes, Rabi flopping, and for sideband scans. Sequence (b) was used to demonstrate suppression of the heating of the WCR modes during simulated clock operation, as displayed in Fig. 5 in the main manuscript, with the central clock operation section repeated a variable number of times in order to measure the evolution of the WCR mode temperature over an increasing number of cycles.}
	\label{fig:sequences}
\end{figure*}

\subsection{Measurements of peak excitation}
In order to improve the efficiency of the measurements, full scans over each line were not performed during measurements of anomalous heating rates (such as those displayed in Fig. \ref{fig:heating}) or during measurements of the Doppler-cooling rate of the WCR modes, but measurements were restricted to the points of peak excitation. To limit any potential systematic error arising from drifts in the WCR mode frequencies during the measurements, the duration of the waiting time to allow the ion to heat was pseudo-randomized, along with whether the measurement was made on the blue or red sideband.

\section{Ground-state cooling of the WCR modes}
\label{sec:gsc}

\subsection{Summary of main approaches to algorithmic cooling}
\label{subsec:approaches}
\begin{figure*}[ht!]
	\centering
	\subfloat[\label{fig:xflops}] {\includegraphics[width=3.4in]{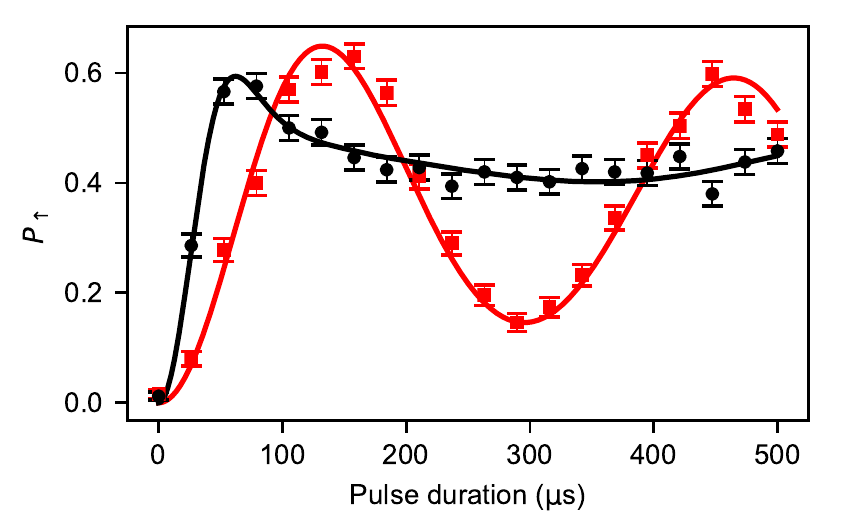} }
	\subfloat[\label{fig:yflops}] {\includegraphics[width=3.4in]{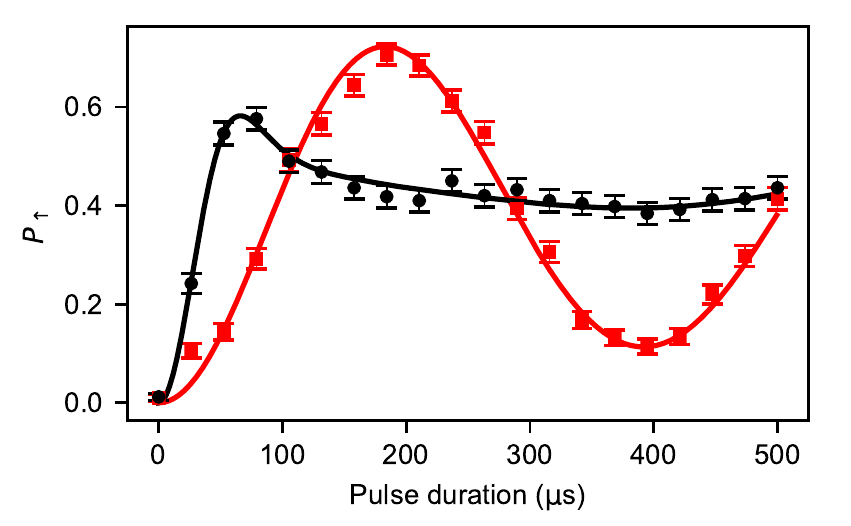} }
	
	\caption{Probability of exciting the HCI to the excited state $P_{\uparrow}(t)$ by driving the first blue sideband of the (a) $x$, and (b) $y$ WCR modes, observed on the HCI as a function of interrogation time before (black) and after (red) algorithmic cooling to the ground state was applied. The clean Rabi flopping behaviour visible after cooling is a clear indication of a high population in the vibrational ground state of both radial in-phase modes. Details of the fitting function is given in the text.}
	\label{fig:flops}
\end{figure*}

Two main approaches are taken for ground-state cooling of the WCR modes. The first is that a separate experimental phase is dedicated to cooling the WCR modes, as shown in Fig. \ref{fig:sequences}(a). This directly follows the optical pumping phase for the HCI, and is composed of eight pulses on the first-order red sideband of each of the two WCR modes, applied in an interleaved fashion to reduce the impact of anomalous heating \cite{leopold_cryogenic_2019}. Owing to the ca. 10~ms lifetime of the excited electronic state of the HCI \cite{lapierre_lifetime_2006, micke_coherent_2020}, after each cooling pulse the HCI must be returned to the ground state using a simplified version of the optical pumping routine presented in reference \cite{micke_coherent_2020}. As the temperatures of the WCR modes decrease during the ground-state cooling process, the optimal pulse duration on these motional sidebands that leads to the highest probability for transfer to the excited state increases gradually \cite{wineland_experimental_1998}. 
This is addressed by gradually increasing the length of the red sideband pulse of the WCR mode between algorithmic cooling cycles. Under typical conditions, an initial value of 90~\micro{s} is used, which is slowly increased to 180~\micro{s}. These values are a compromise between the optimum values for the $x$ and $y$ WCR modes, which are visible in Fig. \ref{fig:flops}. This cooling approach was followed when scanning over the motional sidebands, observing Rabi oscillations (see section~\ref{sec:coherent}), or measuring anomalous heating rates.\par
The second approach to cooling has been employed to produce Fig.~5 of the main text. After initial cooling of the two WCR modes close to their ground states in a procedure similar to the cooling phase in Fig.~\ref{fig:sequences}(a), a sequence resembling optical clock operation is started (see Fig.~\ref{fig:sequences}(b)). During this sequence, only a single algorithmic cooling cycle on each of the two WCR sidebands is added after the HCI electronic state preparation \cite{micke_coherent_2020} during each repetition of the clock sequence. In this case, the pulse time is fixed at 180~\micro{s}. While lacking the raw cooling power of the first approach, this allows an already-cold HCI to be held in the ground state whilst adding minimal experimental overhead.

\subsection{Steady-state temperature after algorithmic cooling}
As mentioned in subsection \ref{subsec:approaches}, the optimum pulse length to achieve the maximum population transfer to the excited state depends on the Fock state $n$. Under our experimental conditions, the Rabi frequencies for the $n=5 \rightarrow n'=4$ transitions of the WCR modes are approximately double that of the $n=1 \rightarrow n'=0$ transitions. This means that there will be no population transfer on the former transition by a pulse whose length matches the so-called $\pi$-time for the latter, leading to difficulty in cooling Fock states higher than $n=5$ to the ground state. Our use of varying pulse lengths during the cooling process mitigates this to some degree, and rough simulations (not including anomalous heating) using our chosen experimental parameters indicate that approximately 80\% of the population reaches the ground state in a given mode, which matches our experimental observations. Further fine-tuning of the pulse lengths, the order in which they are applied, the total number of pulses, and the use of higher-order sidebands to remove multiple phonons per pulse as is done for the axial modes \cite{leopold_cryogenic_2019}, could all potentially increase the ground state populations from the values reached here.

\section{Coherent operations on WCR mode sidebands}
\label{sec:coherent}

The low temperature of the two weakly-coupled radial (WCR) modes after algorithmic cooling can also be observed by the ability to drive coherent oscillations (Rabi flopping) on the blue sidebands, as shown in Fig.~\ref{fig:flops}. Before cooling, no coherent oscillations can be observed due to dephasing from the different Debye-Waller factors \cite{wineland_experimental_1998} of the various Fock states in the initial thermal distribution. After ground-state cooling, the dominant contributor to the signal is the $n=0 \rightarrow n'=1$ transition, leading to a clean oscillation. The maximum observed excitation of approximately 65\% is limited by a combination of imperfect state preparation of the HCI, imperfect quantum logic pulses, and dephasing from the residual Fock state distribution after cooling. The fit to the BSB oscillations on the $x$ WCR mode displayed in Fig.~\ref{fig:flops}(a) takes the form:

\begin{equation}
	P_{\uparrow}(t) = A \sum_m P_{m,n} \sin^{2} (\Omega_{m\rightarrow m+1,n} t)
\end{equation}

where $P_{\uparrow}(t)$ is the probability of finding the ion in the excited state $\ket{\uparrow}$ at time $t$, $A$ is the maximum contrast, $P(m,n)$ is the probability of occupation of Fock states $m$ and $n$ in the $x$ and $y$ WCR modes, respectively, and $\Omega_{m \rightarrow m+1,n}$ is the Rabi frequency for driving state $m,n$ to state $m+1,n$ \cite{wineland_experimental_1998}. For this purpose, an infinite coherence time was assumed. To reduce the number of free parameters in the fit, identical temperatures and thermal distributions in the two modes after cooling were assumed, and any higher-order couplings were neglected. Fock states up to $m, n = 50$ were included in the fit. A similar fit was carried out for the BSB oscillations on the $y$ WCR mode, displayed in Fig.~\ref{fig:flops}(b), but with the appropriate Rabi frequency $\Omega_{m,n \rightarrow n+1}$ and the sum being over $n$. The fitted values of $\langle m \rangle=0.34(3)$, $\langle n \rangle=0.21(2)$ after ground-state cooling are consistent with the values obtained by comparing the peak contrasts of the red and blue sidebands in Fig. 3 of the main manuscript using the procedure described section \ref{sec:nbar}. The values of $\langle m \rangle=6.9(8)$, $\langle n \rangle=7(1)$ extracted for the Doppler-cooled data are higher than those determined from the measurements of the peak contrast, which is likely due to correlations between the fitting parameters in such a complex fit.\par

\section{Anomalous heating}
\label{sec:heating}

\begin{figure}
	\includegraphics[width=3.4in]{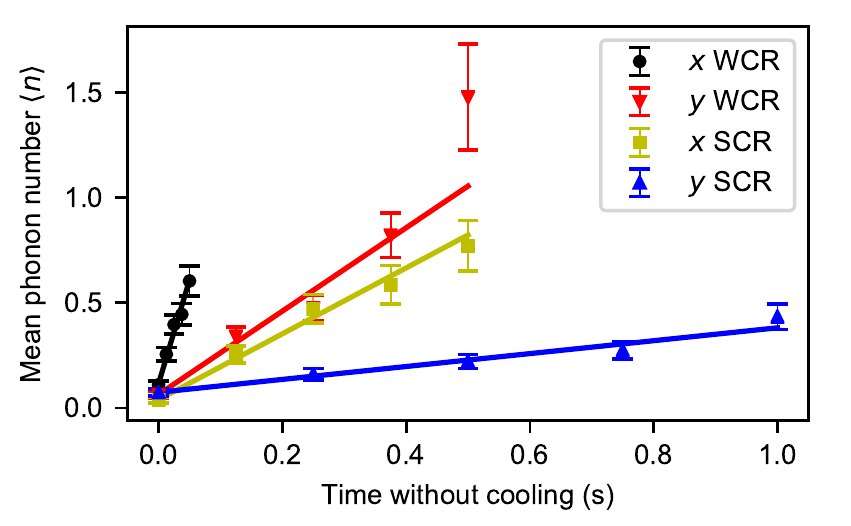}
	\caption{Anomalous heating of the four radial motional modes of the Coulomb crystal: WCR $x$ (black circles), SCR $x$ (red inverted triangles), SCR $y$ (gold squares), WCR $y$ (blue triangles). The fitted linear heating rate (solid lines) is also shown, with the details presented in Table \ref{tab:scrheating}.}
	\label{fig:heating}
\end{figure}

We consider heating of a single ion by electric field noise of a single mode with frequency $\omega$. The heating rate $\Gamma_{h}$ (measured in phonons per second) is related to the single-sided power spectral density of the electric field noise at this frequency $S_{E}(\omega)$ by the following expression: \cite{wineland_experimental_1998-1, brownnutt_ion-trap_2015}

\begin{equation}
	\Gamma_{h} \approx \frac{Z^{2} e^{2}}{4m\hbar\omega} S_{E}(\omega),
\end{equation}
where $e$ is the elementary charge, $\hbar$ the reduced Planck constant, $m$ the mass of the ion, and $Z$ its charge state. The same relation holds in a two-ion crystal in which the heating rate is dominated by one ion, as is the case for the two pairs of radial modes considered here.\par

For measuring these rates, we first cool the two axial modes to the ground state, then the two WCR modes. They are then allowed to heat freely for periods of up to 0.5~seconds. After the delay, we determine the mean number of phonons per mode by measuring their excitation probabilities on the red and blue sidebands using quantum logic (see section \ref{sec:nbar}). To reduce sensitivity to drifts of experimental parameters, the delay times were selected in a pseudo-random sequence. As the quantum logic readout depends crucially on the axial modes staying in the ground state, they are actively cooled during the wait time using stimulated Raman transitions on the Be$^{+}$ ion. During that time, off-resonant effects of the Be$^{+}$ lasers on the WCR modes are negligible owing to the large decoupling, differing mode frequencies, and absence of any other optical transitions from either of the two electronic states of the HCI.\par

In addition to the heating rates of the WCR modes presented in the main manuscript, measurements of the heating rate of the strongly-coupled radial (SCR) out-of-phase modes were also performed in order to provide a cross-check of the earlier results presented in \cite{leopold_cryogenic_2019}. These measurements were performed using Raman transitions in the Be$^{+}$ ion, but the direction of the first Raman beam is inverted compared to the usual orientation as discussed in section \ref{sec:experiment}. This results in the effective projection of the two applied beams being along the radial trap directions, thereby allowing addressing of motional sidebands in this direction.\par

In order to measure the heating rates of the SCR modes, we first cool them both close to their ground states using stimulated Raman transitions, and then allow them to freely heat for a variable time in the absence of all lasers. In this case, ground-state cooling of the axial modes during the wait time is not required as quantum logic operations are unnecessary. The results and the calculated electric field noise power spectral densities are summarized in Fig.~\ref{fig:heating} and Table~\ref{tab:scrheating}, together with the WCR modes for comparison. As discussed in the main manuscript, the observed noise spectral density has improved compared to the data presented in \cite{leopold_cryogenic_2019}, which is most likely attributable to improvements in the electrical supplies for the ion trap that have been implemented since the previous measurements.\par

\begin{table}
\begin{center}
\begin{tabular*}{3.375in}{c @{\extracolsep{\fill}} cccc}

 Axis & Mode & $\omega/2\pi$  & $\Gamma_{h}$  & $S_{E}(\omega)$  \\
 & & (MHz) & (quanta/s) & (V$^{2}$m$^{-2}$Hz$^{-1}$)\\
 \hline
 $x$ & IP (WCR) & 4.59 & 9.9(10) & $1.8(2)\times10^{-15}$\\ 
 $x$ & OP (SCR) & 1.27 & 1.6(1) & $3.1(3)\times10^{-15}$\\ 
 $y$ & IP (WCR) & 4.41 & 2.0(2) & $3.5(3)\times10^{-16}$\\ 
 $y$ & OP (SCR) & 1.03 & 0.31(4) & $4.9(7)\times10^{-16}$\\

\end{tabular*}
\caption{\label{tab:scrheating} Properties of the four radial modes of the two-ion crystal: frequency $\omega/2\pi$, heating rate $\Gamma_{h}$, and calculated electric field noise power spectral density at this frequency $S_{E}(\omega)$.}
\end{center}
\end{table}

Similarly to the WCR modes, we observe a significantly lower heating rate along the $y$ direction than along the $x$ direction. Unfortunately, the relative inflexibility of the motional mode frequencies in this experiment does not allow us to investigate the frequency dependence of the noise in detail. Understanding the sources of anomalous heating in ion traps is an ongoing field of research (see, for example, \cite{sedlacek_evidence_2018}), but our observed scaling seems to lie between the frequency-independent electric field noise that is characteristic of fluctuating dipoles on the ion trap surfaces arising from surface contamination \cite{safavi-naini_influence_2013, safavi-naini_microscopic_2011} and the $1/f$ dependence that would arise from a thin layer of lossy dielectric on the electrode surface \cite{kumph_electric-field_2016}.\par

\section{Second-order Doppler shift due to ion temperature}
In directions in which the trapping is afforded by rf fields (in this case, the radial trap directions), the finite motion of the ion after cooling leads to periodic exposure of the ion to the trapping field. This leads to driven motion, referred to as intrinsic micromotion. Under typical trapping conditions, this leads to an increase of a factor of approximately two in the kinetic energy associated with these modes \cite{berkeland_minimization_1998}.\par
The total time-dilation shift associated with the ion motion in the radial modes can therefore be expressed as \cite{ludlow_optical_2015}:
\begin{equation}
\left( \frac{\Delta \nu}{\nu} \right)_{\mathrm{rad}} \approx -\sum_{i} \frac{\left(\langle n_{i} \rangle + \frac{1}{2} \right) \hbar \omega_{i}}{mc^{2}},
\end{equation}
with $i \in \mathrm{WCR}~x, \mathrm{WCR}~y$.  A conservative target during operation as an optical atomic clock would be $\langle n_{i} \rangle = 0.5$, in which case a total fractional systematic shift of only $-1 \times 10^{-18}$ on the Ar$^{13+}$ transition resonance frequency would result. This calculation neglects the possible non-thermal state of the HCI after cooling \cite{chen_sympathetic_2017}. Producing more precise values would therefore require careful measurements of the motional state distribution. The SCR modes do not contribute to the shift at a meaningful level as the HCI has only a small amplitude of motion in these modes, along with their ease of cooling using the Be$^{+}$ ion.\par

In addition, the rf field experienced by the ions during the secular motion leads to a second-order Stark shift on the transition resonance frequency \cite{ludlow_optical_2015}. This shift is expected to be greatly suppressed in HCIs owing to their extremely small polarizabilities compared to singly-charged ions \cite{kozlov_hci}.\par


%
%

%

\acknowledgments{The authors would like to thank Erik Benkler and Thomas Legero for their contributions to the frequency stabilization of the HCI spectroscopy laser, and Ludwig Krinner for helpful comments on the manuscript. The project was supported by the Physikalisch-Technische Bundesanstalt, the Max-Planck Society, the Max-Planck–Riken–PTB–Center for Time, Constants and Fundamental Symmetries, and the Deutsche Forschungsgemeinschaft (DFG, German Research Foundation) through SCHM2678/5-1, the collaborative research centres SFB 1225 ISOQUANT and SFB 1227 DQ-mat, and under Germany’s Excellence Strategy – EXC-2123 QuantumFrontiers – 390837967. This project 17FUN07 CC4C has received funding from the EMPIR programme co-financed by the Participating States and from the European Union’s Horizon 2020 research and innovation programme. S.A.K. acknowledges financial support from the Alexander von Humboldt Foundation.}


%